\documentclass[11pt,english]{article}
\usepackage[utf8]{inputenc}
\usepackage{float}
\usepackage{amsmath}
\usepackage{amsthm}
\usepackage{amssymb}
\usepackage{graphicx}

\makeatletter

\providecommand{\tabularnewline}{\\}

\@ifundefined{date}{}{\date{}}
\usepackage{t1enc}
\usepackage[english]{babel}
\usepackage{amsthm}
\usepackage{braket}

\def\cros{\raise1.9pt\hbox{$\scriptscriptstyle
          >$}\!\raise1.5pt\hbox{$\scriptstyle\triangleleft\,$}}

\theoremstyle{definition}\theoremstyle{definition}\theoremstyle{definition}\theoremstyle{definition}\newcommand{\noi}{\vspace{0.1in} \noindent}

\title{\bf Three noncontextual hidden variable models \\ for the Peres-Mermin square}
\author{\textit{Gábor Hofer-Szabó}\thanks{Research Center for the Humanities, Budapest, email: szabo.gabor@btk.mta.hu}}

\usepackage{babel}

\makeatother

\usepackage{babel}
\begin{document}
\maketitle 
\begin{abstract}
I will argue that the Peres-Mermin square does not necessarily rule
out a value-definite (deterministic) noncontextual hidden variable
model if the operators are not given a physical interpretation satisfying
the following two requirements: (i) each operator is uniquely realized
by a single physical measurement; (ii) commuting operators are realized
by simultaneous measurements. To underpin this claim, I will construct
three hidden variable models for three different physical realizations
of the Peres-Mermin square: one violating (i), another violating (ii),
and a third one violating both (i) and (ii). \vspace{0.1in}

\end{abstract}

\section{Introduction}

A value-definite (deterministic) hidden variable model for quantum
mechanics is \emph{noncontextual} if the outcome of no measurement
represented by an operator in the formalism depends in any hidden
state on which other simultaneous measurements are performed. The
general aim of this paper is to show that the Kochen-Specker theorems
cannot be successful in ruling out value-definite noncontextual hidden
variable models for quantum mechanics until they are given a physical
interpretation (realization) satisfying the following two requirements: 
\begin{itemize}
\item[(i)] each operator is uniquely realized by a single physical measurement; 
\item[(ii)] commuting operators are realized by simultaneous measurements. 
\end{itemize}
However strange it may seem, it is not at all trivial for a physical
interpretation to satisfy both (i) and (ii). If one realizes each
operator by a different measurement, then some commuting operators
will turn out to be not simultaneously measurable. If one realizes
each set of commuting operators by simultaneous measurements, then
some operators will be realized by multiple measurements. After studying
various interpretations of numerous Kochen-Specker theorems, I have
to admit that I found no such interpretation for any Kochen-Specker
theorem which would satisfy both (i) and (ii). In this paper, however,
I will not argue for the claim that there is no such interpretation
and consequently for the strong claim that there is no valid (algebraic)
argument proving quantum contextuality. (For that see Hofer-Szabó,
2021a, b). Here I will only argue that if an interpretation of a Kochen-Specker
theorem violates either (i) or (ii), then it can be given a value-definite
noncontextual hidden variable model. To make my point, I will pick
a specific Kochen-Specker theorem, the Peres-Mermin square (Peres,
1990; Mermin, 1992) and construct three different noncontextual hidden
variable models for three different physical realizations thereof:
the first violating (i), the second violating (ii), and the third
violating both (i) and (ii). These specific models will simply highlight
the truism that no mathematical no-go theorem \textit{per se} can
prove anything about the physical world without an appropriate physical
interpretation.

In Section 2, I introduce the Peres-Mermin square; in Section 3, I
give three different physical realizations for it; and in Section
4, I construct the three value-definite noncontextual hidden variable
models. I discuss the results in Section 5.

\section{The Peres-Mermin square}

\label{Sec1}

The Peres-Mermin square is the following $3\!\times\!3$ matrix of
self-adjoint operators: 
\begin{center}
\begin{tabular}{ccc}
\quad{}$\sigma_{z}\otimes I$ \quad{}  & \quad{}$I\otimes\sigma_{z}$ \quad{}  & \quad{}$\sigma_{z}\otimes\sigma_{z}$ \quad{}\tabularnewline
 &  & \tabularnewline
\quad{}$I\otimes\sigma_{x}$ \quad{}  & \quad{}$\sigma_{x}\otimes I$ \quad{}  & \quad{}$\sigma_{x}\otimes\sigma_{x}$ \quad{}\tabularnewline
 &  & \tabularnewline
\quad{}$\sigma_{z}\otimes\sigma_{x}$ \quad{}  & \quad{}$\sigma_{x}\otimes\sigma_{z}$ \quad{}  & \quad{}$\sigma_{y}\otimes\sigma_{y}$ \quad{}\tabularnewline
\end{tabular}
\par\end{center}

where $\sigma_{x}$, $\sigma_{y}$, and $\sigma_{z}$ are the Pauli
operators and $I$ is the unit operator on the two dimensional complex
Hilbert space. Each operator in the matrix has two eigenvalues, $\pm1$,
and are arranged in such a way that two operators are commuting if
and only if they are in the same row or in the same column. For example,
\begin{eqnarray*}
[\sigma_{x}\otimes\sigma_{x},\,\sigma_{y}\otimes\sigma_{y}] & = & \sigma_{x}\otimes\sigma_{x}\cdot\sigma_{y}\otimes\sigma_{y}-\sigma_{y}\otimes\sigma_{y}\cdot\sigma_{x}\otimes\sigma_{x}\\
 & = & \sigma_{x}\sigma_{y}\otimes\sigma_{x}\sigma_{y}-\sigma_{y}\sigma_{x}\otimes\sigma_{y}\sigma_{x}\\
 & = & i\sigma_{z}\otimes i\sigma_{z}-(-i)\sigma_{z}\otimes(-i)\sigma_{z}\\
 & = & -\sigma_{z}\otimes\sigma_{z}+\sigma_{z}\otimes\sigma_{z}\\
 & = & 0
\end{eqnarray*}
Commuting operators have common eigenstates. Expressed in the computational
basis 
\begin{eqnarray*}
\ket{0}=\begin{pmatrix}1\\
0
\end{pmatrix}\quad\quad\quad\ket{1}=\begin{pmatrix}0\\
1
\end{pmatrix}
\end{eqnarray*}
and using the notation 
\begin{eqnarray*}
\ket{+}=\frac{1}{\sqrt{2}}\big(\ket{0}+\ket{1}\big)\quad\quad\quad\ket{-}=\frac{1}{\sqrt{2}}\big(\ket{0}-\ket{1}\big)
\end{eqnarray*}
the common eigenstates and the associated eigenvalues of the three
operators in the three subsequent \textit{rows} of the Peres-Mermin
square are the following: 
\begin{table}[H]
\begin{centering}
{\footnotesize{}{}}%
\begin{tabular}{c|ccc}
{\footnotesize{}{}}%
\mbox{%
{\footnotesize{}{}First row}%
}{\footnotesize{}{} }  & {\footnotesize{}{}$\sigma_{z}\otimes I$ }  & {\footnotesize{}{}$I\otimes\sigma_{z}$ }  & {\footnotesize{}{}$\sigma_{z}\otimes\sigma_{z}$}\tabularnewline
\hline 
{\footnotesize{}{}$\ket{\Psi_{1}}=\ket{00}$ }  & {\footnotesize{}{}$+1$ }  & {\footnotesize{}{}$+1$ }  & {\footnotesize{}{}$+1$ }\tabularnewline
{\footnotesize{}{}$\ket{\Psi_{2}}=\ket{01}$ }  & {\footnotesize{}{}$+1$ }  & {\footnotesize{}{}$-1$ }  & {\footnotesize{}{}$-1$ }\tabularnewline
{\footnotesize{}{}$\ket{\Psi_{3}}=\ket{10}$ }  & {\footnotesize{}{}$-1$ }  & {\footnotesize{}{}$+1$ }  & {\footnotesize{}{}$-1$ }\tabularnewline
{\footnotesize{}{}$\ket{\Psi_{4}}=\ket{11}$ }  & {\footnotesize{}{}$-1$ }  & {\footnotesize{}{}$-1$ }  & {\footnotesize{}{}$+1$ }\tabularnewline
\end{tabular}{\footnotesize{}{}\vspace{0.3in}
 }{\footnotesize\par}
\par\end{centering}
\begin{centering}
 
\par\end{centering}
\begin{centering}
{\footnotesize{}{}}%
\begin{tabular}{c|ccc}
{\footnotesize{}{}}%
\mbox{%
{\footnotesize{}{}Second row}%
}{\footnotesize{}{} }  & {\footnotesize{}{}$I\otimes\sigma_{x}$ }  & {\footnotesize{}{}$\sigma_{x}\otimes I$ }  & {\footnotesize{}{}$\sigma_{x}\otimes\sigma_{x}$}\tabularnewline
\hline 
{\footnotesize{}{}$\ket{\Psi'_{1}}=\ket{++}$ }  & {\footnotesize{}{}$+1$ }  & {\footnotesize{}{}$+1$ }  & {\footnotesize{}{}$+1$ }\tabularnewline
{\footnotesize{}{}$\ket{\Psi'_{2}}=\ket{-+}$ }  & {\footnotesize{}{}$+1$ }  & {\footnotesize{}{}$-1$ }  & {\footnotesize{}{}$-1$ }\tabularnewline
{\footnotesize{}{}$\ket{\Psi'_{3}}=\ket{+-}$ }  & {\footnotesize{}{}$-1$ }  & {\footnotesize{}{}$+1$ }  & {\footnotesize{}{}$-1$ }\tabularnewline
{\footnotesize{}{}$\ket{\Psi'_{4}}=\ket{--}$ }  & {\footnotesize{}{}$-1$ }  & {\footnotesize{}{}$-1$ }  & {\footnotesize{}{}$+1$ }\tabularnewline
\end{tabular}{\footnotesize{}{}\vspace{0.3in}
 }{\footnotesize\par}
\par\end{centering}
\begin{centering}
 
\par\end{centering}
\begin{centering}
{\footnotesize{}{}}%
\begin{tabular}{c|ccc}
{\footnotesize{}{}}%
\mbox{%
{\footnotesize{}{}Third row}%
}{\footnotesize{}{} }  & {\footnotesize{}{}$\sigma_{z}\otimes\sigma_{x}$ }  & {\footnotesize{}{}$\sigma_{x}\otimes\sigma_{z}$ }  & {\footnotesize{}{}$\sigma_{y}\otimes\sigma_{y}$}\tabularnewline
\hline 
{\footnotesize{}{}$\ket{\Psi''_{1}}=\frac{1}{\sqrt{2}}\big(\ket{0+}+\ket{1-}\big)$
}  & {\footnotesize{}{}$+1$ }  & {\footnotesize{}{}$+1$ }  & {\footnotesize{}{}$+1$ }\tabularnewline
{\footnotesize{}{}$\ket{\Psi''_{2}}=\frac{1}{\sqrt{2}}\big(\ket{0+}-\ket{1-}\big)$
}  & {\footnotesize{}{}$+1$ }  & {\footnotesize{}{}$-1$ }  & {\footnotesize{}{}$-1$ }\tabularnewline
{\footnotesize{}{}$\ket{\Psi''_{3}}=\frac{1}{\sqrt{2}}\big(\ket{1+}+\ket{0-}\big)$
}  & {\footnotesize{}{}$-1$ }  & {\footnotesize{}{}$+1$ }  & {\footnotesize{}{}$-1$ }\tabularnewline
{\footnotesize{}{}$\ket{\Psi''_{4}}=\frac{1}{\sqrt{2}}\big(\ket{1+}-\ket{0-}\big)$
}  & {\footnotesize{}{}$-1$ }  & {\footnotesize{}{}$-1$ }  & {\footnotesize{}{}$+1$ }\tabularnewline
\end{tabular}
\par\end{centering}
\caption{Common eigenstates and eigenvalues of the operators in the three rows
of the Peres-Mermin square}
\label{table_r} 
\end{table}

Similarly, the common eigenstates and eigenvalues of the three operators
in the three \textit{columns} are the following: 
\begin{table}[H]
\begin{centering}
{\footnotesize{}{}}%
\begin{tabular}{c|ccc}
{\footnotesize{}{}}%
\mbox{%
{\footnotesize{}{}First column}%
}{\footnotesize{}{} }  & {\footnotesize{}{}$\sigma_{z}\otimes I$ }  & {\footnotesize{}{}$I\otimes\sigma_{x}$ }  & {\footnotesize{}{}$\sigma_{z}\otimes\sigma_{x}$}\tabularnewline
\hline 
{\footnotesize{}{}$\ket{\Phi_{1}}=\ket{0+}$ }  & {\footnotesize{}{}$+1$ }  & {\footnotesize{}{}$+1$ }  & {\footnotesize{}{}$+1$ }\tabularnewline
{\footnotesize{}{}$\ket{\Phi_{2}}=\ket{0-}$ }  & {\footnotesize{}{}$+1$ }  & {\footnotesize{}{}$-1$ }  & {\footnotesize{}{}$-1$ }\tabularnewline
{\footnotesize{}{}$\ket{\Phi_{3}}=\ket{1+}$ }  & {\footnotesize{}{}$-1$ }  & {\footnotesize{}{}$+1$ }  & {\footnotesize{}{}$-1$ }\tabularnewline
{\footnotesize{}{}$\ket{\Phi_{4}}=\ket{1-}$ }  & {\footnotesize{}{}$-1$ }  & {\footnotesize{}{}$-1$ }  & {\footnotesize{}{}$+1$ }\tabularnewline
\end{tabular}{\footnotesize{}{}\vspace{0.3in}
 }{\footnotesize\par}
\par\end{centering}
\begin{centering}
 
\par\end{centering}
\begin{centering}
{\footnotesize{}{}}%
\begin{tabular}{c|ccc}
{\footnotesize{}{}}%
\mbox{%
{\footnotesize{}{}Second column}%
}{\footnotesize{}{} }  & {\footnotesize{}{}$I\otimes\sigma_{z}$ }  & {\footnotesize{}{}$\sigma_{x}\otimes I$ }  & {\footnotesize{}{}$\sigma_{x}\otimes\sigma_{z}$}\tabularnewline
\hline 
{\footnotesize{}{}$\ket{\Phi'_{1}}=\ket{+0}$ }  & {\footnotesize{}{}$+1$ }  & {\footnotesize{}{}$+1$ }  & {\footnotesize{}{}$+1$ }\tabularnewline
{\footnotesize{}{}$\ket{\Phi'_{2}}=\ket{-0}$ }  & {\footnotesize{}{}$+1$ }  & {\footnotesize{}{}$-1$ }  & {\footnotesize{}{}$-1$ }\tabularnewline
{\footnotesize{}{}$\ket{\Phi'_{3}}=\ket{+1}$ }  & {\footnotesize{}{}$-1$ }  & {\footnotesize{}{}$+1$ }  & {\footnotesize{}{}$-1$ }\tabularnewline
{\footnotesize{}{}$\ket{\Phi'_{4}}=\ket{-1}$ }  & {\footnotesize{}{}$-1$ }  & {\footnotesize{}{}$-1$ }  & {\footnotesize{}{}$+1$ }\tabularnewline
\end{tabular}{\footnotesize{}{}\vspace{0.3in}
 }{\footnotesize\par}
\par\end{centering}
\begin{centering}
 
\par\end{centering}
\begin{centering}
{\footnotesize{}{}}%
\begin{tabular}{c|ccc}
{\footnotesize{}{}}%
\mbox{%
{\footnotesize{}{}Third column}%
}{\footnotesize{}{} }  & {\footnotesize{}{}$\sigma_{z}\otimes\sigma_{z}$ }  & {\footnotesize{}{}$\sigma_{x}\otimes\sigma_{x}$ }  & {\footnotesize{}{}$\sigma_{y}\otimes\sigma_{y}$}\tabularnewline
\hline 
{\footnotesize{}{}$\ket{\Phi''_{1}}=\frac{1}{\sqrt{2}}\big(\ket{00}+\ket{11}\big)$
}  & {\footnotesize{}{}$+1$ }  & {\footnotesize{}{}$+1$ }  & {\footnotesize{}{}$-1$ }\tabularnewline
{\footnotesize{}{}$\ket{\Phi''_{2}}=\frac{1}{\sqrt{2}}\big(\ket{00}-\ket{11}\big)$
}  & {\footnotesize{}{}$+1$ }  & {\footnotesize{}{}$-1$ }  & {\footnotesize{}{}$+1$ }\tabularnewline
{\footnotesize{}{}$\ket{\Phi''_{3}}=\frac{1}{\sqrt{2}}\big(\ket{01}+\ket{10}\big)$
}  & {\footnotesize{}{}$-1$ }  & {\footnotesize{}{}$+1$ }  & {\footnotesize{}{}$+1$ }\tabularnewline
{\footnotesize{}{}$\ket{\Phi''_{4}}=\frac{1}{\sqrt{2}}\big(\ket{01}-\ket{10}\big)$
}  & {\footnotesize{}{}$-1$ }  & {\footnotesize{}{}$-1$ }  & {\footnotesize{}{}$-1$ }\tabularnewline
\end{tabular}
\par\end{centering}
\caption{Common eigenstates and eigenvalues of the operators in the three columns
of the Peres-Mermin square}
\label{table_c} 
\end{table}

\noi Observe that the common eigenstates of the operators in the
first two rows and columns are product states, whereas those in the
third row and column are entangled states. The vectors in the third
column form the so-called Bell basis with $\ket{\Phi''_{4}}$ as the
singlet state. However, I will also refer to the vectors in the third
row as Bell states.

Also observe that the product of the eigenvalues of the three operators
in all the three rows and in the first two columns is $+1$, whereas
it is $-1$ for the operators in the third column. As a mathematical
consequence, one cannot fill in a $3\!\times\!3$ matrix with numbers
$\pm1$ such that these numbers conform in each row and column to
one of the four triples of eigenvalues. Call this mathematical fact
the \emph{Peres-Mermin contradiction}. The Peres-Mermin contradiction
is often interpreted as the impossibility to provide a value-definite
noncontextual hidden variable model for quantum mechanics. Namely,
such a model should consist of such hidden states which assign values
to the operators in such a way that any triple of operators in a row
or column have values according to one of the four triples of eigenvalues
in that row or column.

Note, however, that without a physical realization of the operators
such a conclusion is without any physical justification. And indeed,
in the next Section we will provide three different physical realizations
of the Peres-Mermin square for which there \textit{is} a value-definite
noncontextual hidden variable model. The models for these realizations
will avoid the Peres-Mermin contradiction in the following two ways.
First, if the realization violates requirement (i)---that is some
operators are realized by multiple measurements, then the model will
not assing the same value for these measurements in some of the hidden
states. (Note that the fact that physically different measurements
are represented by the same operator in quantum mechanics does not
mean that they need to have the some outcome in every hidden state.
All that is required is that they have the same \emph{distribution}
of outcomes in every quantum state.) Second, if the realization violates
requirement (ii)---that is some commuting operators are not realized
by simultaneous measurements, then the model will not respect the
constraint coming from the corresponding triples of eigenvalues since
this constraint is not empirically justified. (Note that for the empirical
justification of the constraint that in a certain row or column the
admissible triples of eigenvalues are just those four which are in
Table \ref{table_r} or \ref{table_c}, we need to perform the three
measurements realizing the corresponding three operators \emph{simultanously}
and check which combinations of outcomes come up and which do not.)
Thus, one can avoid the Peres-Mermin contradiction and construct a
value-definite noncontextual hidden variable model for a given realization
of the Peres-Mermin square if either (i) or (ii) is violated. In the
first case one avoids the contradiction by writting more than one
number in some entries of the matrix; in the second case by filling
in the matrix such that it conforms only to those constraints in Table
\ref{table_r} or \ref{table_c} which are backed by simultaneous
measurements. 

Now, let us turn to the concrete realizations. 

\section{Three realizations of the Peres-Mermin square}

A realization (interpretation) of the Peres-Mermin square is an association
of operators in the matrix with real-world measurements. In all three
realizations to be provided, the operators will represent measurements
performed on pairs of photons prepared in a given quantum state. The
realizations will differ in what these measurements exactly are.

\noi \textit{First realization}. The schematic picture of the first
realization is portrayed in Figure \ref{Fig1}: 
\begin{figure}[H]
\centerline{\resizebox{8cm}{!}{\includegraphics{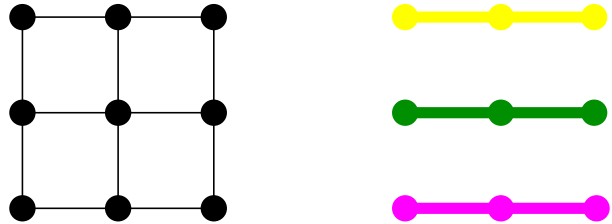}}} \caption{First realization of the Peres-Mermin square}
\label{Fig1} 
\end{figure}

On the left graph, the black vertices represent the 9 operators in
the Peres-Mermin square and the \big(hyper)edges connecting the vertices
represent the commutativity relation between the operators. On the
right graph, the coloured vertices represent the 9 measurements uniquely
associated with the operators and the edges represent the simultaneous
measurability relation.

What are these 9 measurements?

Well, actually there are only three physical measurement procedures,
to the result of which one applies three different functions. (This
is why you find only three different colors on the graph each representing
a different physical measurement. The thickened edges express the
fact that the measurements in each triple are \textit{trivially} simultaneously
measurable: they are just different functions of the same physical
measurement.) The three physical measurements are the following: 
\begin{description}
\item [{$L_{zz}$:}] measure the \emph{linear polarization} of both photons
along a given transverse axis $z$ (with outcome $+1$ if the photon
passes the polarizer and $-1$ if not); 
\item [{$L_{xx}$:}] measure the \emph{linear polarization} of both photons
along an axis $x$ at $45^{\circ}$ from the axis $z$; 
\item [{$B$:}] perform a \emph{Bell state measurement} on the photon pair
with four outcomes corresponding to the Bell states $\ket{\Psi''_{1}}$,
$\ket{\Psi''_{2}}$, $\ket{\Psi''_{3}}$, and $\ket{\Psi''_{4}}$
in Table \ref{table_r}. 
\end{description}
The 9 measurements arise from the 3 physical measurements by applying
certain functions on the measurement results. After performing the
linear polarization measurement $L_{zz}$ and obtaining four different
combinations of outcomes ($\pm1$ on the left and $\pm1$ on the right),
one can do three different things: either one registers only the outcome
on the \emph{left} wing ($l(L_{zz})$); or the outcome on the \emph{right}
wing ($r(L_{zz})$); or registers the \emph{product} of the two outcomes
($t(L_{zz})$). Similarly, one can define $l(L_{xx})$, $r(L_{xx})$,
and $t(L_{xx})$ from applying the same functions to the result of
$L_{xx}$. Finally, one can apply three functions $f$, $g$, or $h$
to the results of the Bell state measurement $B$. These functions
assign to each outcome of $B$ the corresponding eigenvalue of the
operators $\sigma_{z}\otimes\sigma_{x}$, $\sigma_{x}\otimes\sigma_{z}$,
and $\sigma_{y}\otimes\sigma_{y}$ according to the third row of Table
\ref{table_r}. For example, if the outcome of the Bell state measurement
$B$ is $\ket{\Psi''_{1}}$, then $f(B)=+1$, $g(B)=+1$ and $h(B)=+1$.

The operators and the corresponding measurements realizing the operators
are portrayed in Table \ref{table3}: 
\begin{table}[H]
\begin{centering}
\begin{tabular}{ccc|ccc}
 & Operators  &  &  & Measurements  & \tabularnewline
\hline 
\hline 
 &  &  &  &  & \tabularnewline
\quad{}$\sigma_{z}\otimes I$ \quad{}  & \quad{}$I\otimes\sigma_{z}$ \quad{}  & \quad{}$\sigma_{z}\otimes\sigma_{z}$ \quad{}  & \quad{}$l(L_{zz})$  & \quad{}$r(L_{zz})$  & \quad{}$t(L_{zz})$ \quad{}\tabularnewline
 &  &  &  &  & \tabularnewline
\quad{}$I\otimes\sigma_{x}$ \quad{}  & \quad{}$\sigma_{x}\otimes I$ \quad{}  & \quad{}$\sigma_{x}\otimes\sigma_{x}$ \quad{}  & \quad{}$r(L_{xx})$  & \quad{}$l(L_{xx})$  & \quad{}$t(L_{xx})$ \quad{}\tabularnewline
 &  &  &  &  & \tabularnewline
\quad{}$\sigma_{z}\otimes\sigma_{x}$ \quad{}  & \quad{}$\sigma_{x}\otimes\sigma_{z}$ \quad{}  & \quad{}$\sigma_{y}\otimes\sigma_{y}$ \quad{}  & \quad{}$f(B)$  & \quad{}$g(B)$  & \quad{}$h(B)$ \quad{}\tabularnewline
\end{tabular}
\par\end{centering}
\caption{First realization of the Peres-Mermin square}
\label{table3} 
\end{table}

Now, the realization satisfies requirement (i) since each operator
is uniquely realized by a measurement. But it violates requirement
(ii) since the measurements, for example, realizing the commuting
operators in the third column $t(L_{zz})$, $t(L_{xx})$, and $h(B)$
are \textit{not simultaneously measurable} for the simple reason that
no two of the three measurements $L_{zz}$, $L_{xx}$ and $B$ can
be performed on the same pair of photons at the same time. Note that
the measurement $L_{zz}$ requires two polarization beam splitters
each oriented along direction $z$ on the opposite wings; the measurement
$L_{xx}$ requires two polarization beam splitters each oriented along
direction $x$ on the opposite wings; and the Bell state measurement
$B$ requires a complicated arrangement of beam splitters and polarization
beam splitters in the way of the two photons (Lütkenhaus et al., 1999).
These measurement arrangements are incompatible, they cannot be performed
simultaneously on the same pair of photons. Consequently, one cannot
experimentally verify via these measurements whether the outcomes
of $t(L_{zz})$, $t(L_{xx})$, and $h(B)$ \textit{in a given run
of the experiment} conform to one of the rows of Table \ref{table_c}
or not. But then neither the hidden variable models need to respect
these constraints. We came back to this point in Section \ref{Sec3}.

Now, one might wonder whether the lack of simultaneous measurability
between the measurement represented by the vertical triples of operators
can be cured by introducing new measurements. This leads us to the
second realization of the Peres-Mermin square.

\noi \textit{Second realization}. The second realization is portrayed
in Figure \ref{Fig2}: 
\begin{figure}[H]
\centerline{\resizebox{8cm}{!}{\includegraphics{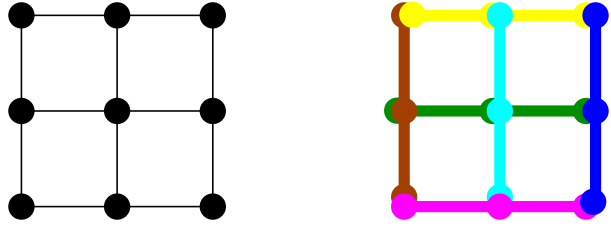}}} \caption{Second realization of the Peres-Mermin square}
\label{Fig2} 
\end{figure}

As we see on the right graph, we have now 6 measurements: the three
old measurements: $L_{zz}$, $L_{xx}$ and $B$ plus three new ones
realizing the three vertical triples: 
\begin{description}
\item [{$L_{zy}$:}] measure the \emph{linear polarization} of the left
photon along the axis $z$ and the linear polarization of the right
photon along the axis $x$; 
\item [{$L_{xz}$:}] measure the \emph{linear polarization} of the left
photon along the axis $x$ and the linear polarization of the right
photon along the axis $z$; 
\item [{$B'$:}] perform another \emph{Bell state measurement} on the photon
pair with four outcomes corresponding to the Bell states $\ket{\Phi''_{1}}$,
$\ket{\Phi''_{2}}$, $\ket{\Phi''_{3}}$, and $\ket{\Phi''_{4}}$
in Table \ref{table_c}. 
\end{description}
Just as above, there are 9 measurements arising from the 3 new physical
measurements: $l(L_{zx})$, $r(L_{zx})$, $t(L_{zx}$); and $l(L_{xz})$,
$r(L_{xz})$, $t(L_{xz}$); and finally $f'(B')$, $g'(B')$, $h'(B')$.
(In these latter measurement, we again apply three functions $f'$,
$g'$ and $h'$ assigning to each outcome of the Bell state measurement
$B'$ the corresponding eigenvalue of the operators $\sigma_{z}\otimes\sigma_{z}$,
$\sigma_{x}\otimes\sigma_{x}$, and $\sigma_{y}\otimes\sigma_{y}$
according to the third column of Table \ref{table_c}.) The operators
and corresponding measurements realizing the operators are portrayed
in Table \ref{table4}: 
\begin{table}[H]
\begin{centering}
\begin{tabular}{ccc|ccc}
 & Operators  &  &  & Measurements  & \tabularnewline
\hline 
\hline 
 &  &  &  &  & \tabularnewline
\quad{}$\sigma_{z}\otimes I$ \quad{}  & \quad{}$I\otimes\sigma_{z}$ \quad{}  & \quad{}$\sigma_{z}\otimes\sigma_{z}$ \quad{}  & \quad{}$l(L_{zz}$ / $l(L_{zx})$  & \quad{}$r(L_{zz})$ / $r(L_{xz})$  & \quad{}$t(L_{zz})$ / $f'(B')$ \quad{}\tabularnewline
 &  &  &  &  & \tabularnewline
\quad{}$I\otimes\sigma_{x}$ \quad{}  & \quad{}$\sigma_{x}\otimes I$ \quad{}  & \quad{}$\sigma_{x}\otimes\sigma_{x}$ \quad{}  & \quad{}$r(L_{xx})$ / $r(L_{zx})$  & \quad{}$l(L_{xx})$ / $l(L_{xz})$  & \quad{}$t(L_{xx})$ / $g'(B')$ \quad{}\tabularnewline
 &  &  &  &  & \tabularnewline
\quad{}$\sigma_{z}\otimes\sigma_{x}$ \quad{}  & \quad{}$\sigma_{x}\otimes\sigma_{z}$ \quad{}  & \quad{}$\sigma_{y}\otimes\sigma_{y}$ \quad{}  & \quad{}$f(B)$ / $t(L_{zx})$  & \quad{}$g(B)$ / $t(L_{xz})$  & \quad{}$h(B)$ / $h'(B')$ \quad{}\tabularnewline
\end{tabular}
\par\end{centering}
\caption{Second realization of the Peres-Mermin square}
\label{table4} 
\end{table}

Now, this realization of the Peres-Mermin square, contrary to the
first realization, satisfies requirement (ii) since commuting triples
of operators are all realized by simultaneous measurements. However,
requirement (i) is now violated since every operator is realized by
two different measurements. Four of the nine realizations are unproblematic:
due to locality, the measurements $l(L_{zz})$ and $l(L_{zx})$ can
be considered the same measurements since in both measurements one
does the same thing in the left wing of the experiment. Similarly,
one can argue that 
\begin{eqnarray*}
r(L_{zz})=r(L_{xz})\quad\quad r(L_{xx})=r(L_{zx})\quad\quad l(L_{xx})=l(L_{xz})
\end{eqnarray*}
However, the other five pairs of measurement are not the same measurements:
\begin{eqnarray*}
t(L_{zz})\neq f'(B')\quad\quad t(L_{xx})\neq g'(B')\quad\quad f(B)\neq t(L_{zx})\quad\quad g(B)\neq t(L_{xz})\quad\quad h(B)\neq h'(B')
\end{eqnarray*}
For example, $t(L\_\{zz\})$ and $f'(B')$, both realizing the operator
$\sigma_{z}\otimes\sigma_{z}$ in the upper right entry, are two physically
different measurement arrangements; they cannot be performed at the
same time on the same pair of photons. Thus, requirement (i) is violated.

Now, let us go over to the third realization where neither (i) nor
(ii) holds.

\noi \textit{Third realization}. The third realization is portrayed
in Figure \ref{Fig3}: 
\begin{figure}[H]
\centerline{\resizebox{8cm}{!}{\includegraphics{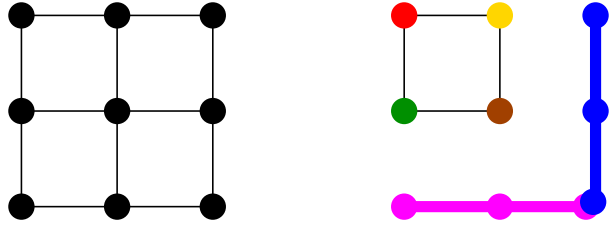}}} \caption{Third realization of the Peres-Mermin square}
\label{Fig3} 
\end{figure}

As we can see on the right graph, we have now again 6 measurements: 
\begin{description}
\item [{$L_{l(z)}$:}] measure the linear polarization of the left photon
along the axis $z$; 
\item [{$L_{r(z)}$:}] measure the linear polarization of the right photon
along the axis $z$; 
\item [{$L_{l(x)}$:}] measure the linear polarization of the left photon
along the axis $x$; 
\item [{$L_{r(x)}$:}] measure the linear polarization of the right photon
along the axis $x$; 
\item [{$B$:}] perform a Bell state measurement with four outcomes corresponding
to the Bell states $\ket{\Psi''_{1}}$, $\ket{\Psi''_{2}}$, $\ket{\Psi''_{3}}$,
and $\ket{\Psi''_{4}}$; 
\item [{$B'$:}] perform a Bell state measurement with four outcomes corresponding
to the Bell states $\ket{\Phi''_{1}}$, $\ket{\Phi''_{2}}$, $\ket{\Phi''_{3}}$,
and $\ket{\Phi''_{4}}$. 
\end{description}
From the two Bell state measurements we again obtain the measurements
$f(B)$, $g(B)$, $h(B)$ and $f'(B')$, $g'(B')$, $h'(B')$. The
operators and corresponding measurements realizing the operators are
portrayed in Table \ref{table5}: 
\begin{table}[H]
\begin{centering}
\begin{tabular}{ccc|ccc}
 & Operators  &  &  & Measurements  & \tabularnewline
\hline 
\hline 
 &  &  &  &  & \tabularnewline
\quad{}$\sigma_{z}\otimes I$ \quad{}  & \quad{}$I\otimes\sigma_{z}$ \quad{}  & \quad{}$\sigma_{z}\otimes\sigma_{z}$ \quad{}  & \quad{}$L_{l(z)}$  & \quad{}$L_{r(z)}$  & \quad{}$f'(B')$ \quad{}\tabularnewline
 &  &  &  &  & \tabularnewline
\quad{}$I\otimes\sigma_{x}$ \quad{}  & \quad{}$\sigma_{x}\otimes I$ \quad{}  & \quad{}$\sigma_{x}\otimes\sigma_{x}$ \quad{}  & \quad{}$L_{r(x)}$  & \quad{}$L_{l(x)}$  & \quad{}$g'(B')$ \quad{}\tabularnewline
 &  &  &  &  & \tabularnewline
\quad{}$\sigma_{z}\otimes\sigma_{x}$ \quad{}  & \quad{}$\sigma_{x}\otimes\sigma_{z}$ \quad{}  & \quad{}$\sigma_{y}\otimes\sigma_{y}$ \quad{}  & \quad{}$f(B)$  & \quad{}$g(B)$  & \quad{}$h(B)$ / $h'(B')$ \quad{}\tabularnewline
\end{tabular}
\par\end{centering}
\caption{Third realization of the Peres-Mermin square}
\label{table5} 
\end{table}

Now, in this third realization of the Peres-Mermin square both requirement
(i) and (ii) are violated: the operator $\sigma_{y}\otimes\sigma_{y}$
in the bottom right corner is realized by two different measurements
$h(B)$ and $h'(B')$; and the measurements $L_{r(z)}$ and $f'(B')$
are not simultaneously measurable (since the linear polarization measurement
and the Bell state measurement require different experimental arrangements;
see again Lütkenhaus et al., 1999).

\noi To sum up, we have three different realizations of the Peres-Mermin
square: one violating requirement (i), one violating requirement (ii),
and a third one violating both (i) and (ii). Next, we construct a
value-definite noncontextual hidden variable model for each realization.

\section{Three value-definite noncontextual hidden variable models for the
Peres-Mermin square}

\label{Sec3}

\textit{First realization}. Let $\ket{\Psi}$ denote the quantum state
of the beam of the photon pairs. A value-definite noncontextual hidden
variable model for the first realization of the Peres-Mermin square
is the following. We have three physical measurements, $L_{zz}$,
$L_{_{xx}}$, and $B$, each with four outcomes. Let $i,j,k=1\dots4$
denote the indices running through these outcomes. The hidden variable
model will then consist of $4^{3}$ deterministic hidden states 
\begin{eqnarray*}
\lambda^{ijk}\quad\quad\quad\quad\quad\quad i,j,k=1\dots4
\end{eqnarray*}
providing the outcomes $i$, $j$ and $k$ for the measurements $L_{zz}$,
$L_{xx}$, and $B$, respectively. The probability of these hidden
states is 
\begin{eqnarray*}
p(\lambda^{ijk})=|\!\braket{\Psi_{i},\Psi}\!|^{2}\cdot|\!\braket{\Psi'_{j},\Psi}\!|^{2}\cdot|\!\braket{\Psi''_{k},\Psi}\!|^{2}
\end{eqnarray*}
Clearly, these probabilities add up to 1 and the model recovers the
statistical predictions of quantum mechanics with respect to the above
three measurements.

Note that this hidden variable model does not respect Table \ref{table_c}
in Section \ref{Sec1}. In other words, it is possible that some hidden
states (with nonzero probability) will provide such outcomes for a
triple of measurements in a given \textit{column} which do not conform
to any triple of eigenvalues in Table \ref{table_c}. For example,
for the measurements $t(L)$, $t(L)$, and $h(B)$ represented by
the operators $\sigma_{z}\otimes\sigma_{z}$, $\sigma_{x}\otimes\sigma_{x}$,
and $\sigma_{y}\otimes\sigma_{y}$ in the third column, the hidden
state $\lambda^{111}$ will provide the outcomes 
\begin{eqnarray*}
t(L_{zz})=+1\quad\quad\quad t(L_{xx})=+1\quad\quad\quad h(B)=+1
\end{eqnarray*}
The triple $(+1,+1,+1)$, however, is not an admissible triple of
eigenvalues according to the third column of Table \ref{table_c}.
But this is no problem since the measurements $L_{zz}$, $L_{xx}$,
and $B$, and consequently $t(L)$, $t(L)$, and $h(B)$ cannot be
performed simultaneously in a given run of the experiment. Thus, the
hidden variable model need not respect Table \ref{table_c}. (Note
that Table \ref{table_c} should be respected by a model only if it
contains simultaneous measurements represented by operators in the
\textit{columns} of the Peres-Mermin square. Otherwise Table \ref{table_c}
does not put any constraint on the model.)

\noi \textit{Second and third realization}. For the second and third
realization of the Peres-Mermin square we construct the same value-definite
noncontextual hidden variable model. In both cases we have six physical
measurements. In the second realization: 
\begin{eqnarray}
L_{zz}\quad\quad L_{xx}\quad\quad L_{zx}\quad\quad L_{xz}\quad\quad B\quad\quad B'\label{seconda}
\end{eqnarray}
with 
\begin{eqnarray}
l(L_{zz})=l(L_{zx})\quad\quad r(L_{zz})=r(L_{xz})\quad\quad r(L_{xx})=r(L_{zx})\quad\quad l(L_{xx})=l(L_{xz})\label{secondb}
\end{eqnarray}
In the third realization: 
\begin{eqnarray}
L_{l(z)}\quad\quad L_{r(z)}\quad\quad L_{l(x)}\quad\quad L_{r(x)}\quad\quad B\quad\quad B'\label{third}
\end{eqnarray}
The measurements (\ref{seconda})-(\ref{secondb}) and (\ref{third}),
however, can easily be translated into one another via Table \ref{table_r}
and \ref{table_c} in a one-to-one way. For example, 
\begin{eqnarray*}
L_{zz}=1\quad\quad L_{xx}=4\quad\quad L_{zx}=2\quad\quad L_{xz}=2
\end{eqnarray*}
if and only if 
\begin{eqnarray*}
L_{l(z)}=+1\quad\quad L_{r(z)}=+1\quad\quad L_{l(x)}=-1\quad\quad L_{r(x)}=-1
\end{eqnarray*}
and similarly for the other 15 measurement outcomes. So it is enough
to provide a hidden variable model for the third realization.

Our hidden variable model will consist of $2^{4}\cdot4^{2}$ deterministic
hidden states 
\begin{eqnarray*}
\lambda^{ijklmn}\quad\quad\quad\quad\quad\quad i,j,k,l=\pm1,\quad m,n=1\dots4
\end{eqnarray*}
providing the outcomes $i\dots n$ for the six measurements in (\ref{third}).
The probability of these hidden states is 
\begin{eqnarray*}
p(\lambda^{ijklmn})=p^{ijkl}\cdot|\!\braket{\Psi''_{m},\Psi}\!|^{2}\cdot|\!\braket{\Phi''_{n},\Psi}\!|^{2}
\end{eqnarray*}
where the probabilities $p^{ijkl}$ can be straightforwardly constructed
using Arthur Fine's method developed in detail in Proposition 2 in
(Fine, 1982). Here we do not repeat the construction; we only note
that since the eight eigenprojections 
\begin{eqnarray}
P_{\sigma_{z}\otimes I}^{\pm}\quad\quad\quad P_{I\otimes\sigma_{z}}^{\pm}\quad\quad\quad P_{\sigma_{x}\otimes I}^{\pm}\quad\quad\quad P_{I\otimes\sigma_{x}}^{\pm}\label{projections}
\end{eqnarray}
of the four operators in the $2\!\times\!2$ top left submatrix of
the Peres-Mermin square do not (in any combination) violate the Clauser-Horne
inequality, there exists a value-definite noncontextual hidden variable
model for any quantum state $\ket{\Psi}$. Fine explicitly constructs
the hidden variable model with the 16 probabilities $p^{ijkl}$. (One
just substitutes the projections with $+1$ eigenvalues in (\ref{projections})
for Fine's $A,B,A',B'$ and the projections with $-1$ eigenvalues
for $\overline{A},\overline{B},\overline{A}',\overline{B}'$, respectively,
to get the probabilities $p^{ijkl}$.) For example, if $\ket{\Psi}=\ket{\Psi_{1}}$,
we obtain 
\begin{eqnarray*}
p^{+1+1kl}=\frac{1}{4}\quad\quad\quad\quad\quad\quad k,l=\pm1
\end{eqnarray*}
and all the other probabilities are $0$.

Note again that this hidden variable model is also not needed to conform
to those constraints coming from Table \ref{table_r} or \ref{table_c}
which are not backed by simultaneous measurability. Consider, for
example, the operator $\sigma_{z}\otimes\sigma_{z}$. On the second
realization, $\sigma_{z}\otimes\sigma_{z}$ is doubly realized: it
can be measured either as $t(L_{zz})$ or as $f'(B')$. Consider now
the hidden state $\lambda^{+1+1klm4}$. In this hidden state $t(L_{zz})=1$
assigning the eigenvalue $+1$ to $\sigma_{z}\otimes\sigma_{z}$ (according
to the first row in Table \ref{table_r}) and $f'(B')=4$ assigning
the eigenvalue $-1$ to $\sigma_{z}\otimes\sigma_{z}$ (according
to the third column in Table \ref{table_c}). But since $t(L)$ and
$f'(B')$ are different measurements, the two eigenvalues (measurement
outcomes) need not match. (Still, one can easily find a $\ket{\Psi}$
for which $p(\lambda^{+1+1klm4})\neq0$.)

Similarly, on the third interpretation, the measurements $P_{r(z)}$
and $f'(B')$ are not simultaneously measurable. Therefore, the hidden
states need not conform to first row of Table \ref{table_r}.

Let me note, finally, that all the above models are noncontextual
since the response of the system to a measurement in a given hidden
state does \textit{not} depend on which other measurements are simultaneously
performed with it. The model also is nonconspiratorial since the distribution
of the hidden states does not depend on the measurement choices (which
are not even mentioned).

\section{Discussion}

In the paper it was shown that the Peres-Mermin square does not necessarily
rule out a noncontextual hidden variable model if the physical realization
of the operators does not satisfy the following two requirements: 
\begin{itemize}
\item[(i)] each of the 9 operators is uniquely realized by a single measurement; 
\item[(ii)] commuting operators are realized by simultaneous measurements. 
\end{itemize}
To make my point, I constructed three hidden variable models for three
different physical realizations of the Peres-Mermin square: one violating
(i), another violating (ii), and a third one violating both (i) and
(ii). These models were not to suggest that quantum mechanics might
admit a noncontextual hidden variable model. Far from it, we know
from the violation of the Bell inequalities that this is impossible.
Rather, by the above constructions I intended to show that in order
to interpret the Kochen-Specker theorems as proving quantum contextuality,
one needs to carefully interpret the operators in the theorems. More
concretely, the Kochen-Specker theorems prove quantum contextuality
only if the operators featuring in the theorem can be given a physical
interpretation satisfying requirements (i) and (ii).

Now, requirement (i) is often ignored because it is implicitly required
from a hidden variable model that any two measurements which are represented
by the same operator should give the same outcome in every hidden
state. Sometimes this constraint is also called noncontextuality (Spekkens,
2005). In (Hofer-Szabó 2021a, b) I argued against calling this constraint
noncontextuality by showing that the two constraints are logically
independent. Note that this constraint is violated in the second and
third realization of the Peres-Mermin square while both models are
noncontextual in the sense defined above.

As for requirement (ii), it is often tacitly assumed that commutativity
and simultaneous measurability are synonyms. But they are not. In
(Hofer-Szabó 2021a, b) again, I argued that from the pure fact that
two operators are commuting it does not follow that any two measurements
realizing the operators can be performed at the same time on the same
system. The two measurement procedures may well be physically incompatibility,
just like the polarization measurement and Bell state measurements
in the second and third realization. But realizing commuting operators
by simultaneous measurements is necessary if we want to physically
justify the mathematical no-go theorems.

In this paper, however, I did not intend to argue for requirement
(i) and (ii). I simply wanted to show that they are \textit{sine qua
non} for taking the Kochen-Specker theorems to prove quantum contextuality.

\vspace{0.2in}
\noi \textbf{Acknowledgements.} This work has been supported by the
Hungarian National Research, Development and Innovation Office, K-115593
and K-134275 and a Senior Research Scholarship of the Institute of
Advanced Studies Koszeg. I wish to thank Thomas Filk for valuable
discussions.

\section*{References}
\begin{description}
\item A. Fine, “Hidden variables, joint probability, and the Bell inequalities” Phys. Rev. Lett. 48, 291–295 (1982).

\item N. Lütkenhaus, J. Calsamiglia, K-A. Suominen, “On Bell measurements for teleportation,” Phys. Rev. A, 59, 3295.

\item G. Hofer-Szabó, ``Commutativity, comeasurability, and contextuality in the Kochen-Specker arguments,'' Phil. Sci., (forthcoming, 2021a).

\item G. Hofer-Szabó, ``Two concepts of noncontextuality in quantum mechanics,'' (submitted, 2021b).

\item D. Mermin, “Ontological states and the two theorems of John Bell,” Rev. Mod. Phys., 65 (3), 803-815 (1993).

\item A. Peres, “Incompatible Results of Quantum Measurements,” Phys. Lett. A, 151, 107-108 (1990).

\item R. W. Spekkens, “Contextuality for preparations, transformations, and unsharp measurements,” Phys. Rev. A, 71, 052108 (2005). 
\end{description}

\end{document}